# Highly organized smectic-like packing in vapor-deposited glasses of a liquid crystal

*Ankit Gujral, Jaritza Gómez, Jing Jiang, Chengbin Huang, Kathryn A. O'Hara, Michael F. Toney, Michael L. Chabinyc, Lian Yu, M.D. Ediger*

## Abstract

Glasses of a model smectic liquid crystal-forming molecule, itraconazole, were prepared by vapor deposition onto substrates with temperatures ranging from $T_{substrate} = 0.78\ T_g$ to $1.02\ T_g$, where $T_g = 330$ K is the glass transition temperature. The films were characterized using x-ray scattering techniques. For $T_{substrate}$ near and below $T_g$, glasses with layered smectic-like structures can be prepared and the layer spacing can be tuned by 16% through choice of $T_{substrate}$. Remarkably, glasses prepared with $T_{substrate}$ above $T_g$ exhibit much higher structural organization than a thermally annealed film. These results are explained by a mechanism based upon preferred molecular orientation and enhanced molecular motion at the free surface, indicating that molecular organization in the glass is independent of the anchoring preferred at the substrate. These results suggest new strategies of optimizing molecular packing within active layers of organic electronic and optoelectronic devices.

# Introduction

Control of molecular organization in solids, a long-standing goal in the chemical and materials sciences, is needed to advance many different areas of science and technology. This issue is at the heart of crystal engineering[1,2] and is essential for many types of organic electronic devices. For example, in organic field effect transistors the selection of crystal polymorph and orientation serves to optimize charge transport.[3] The emitting molecules can be oriented in a glassy host to maximize light output in organic light emitting diodes.[4] For organic photovoltaics, anisotropic amorphous packing can help to maximize π-stacking and charge transport.[5] While some approaches in these fields utilize equilibrium states of matter, metastable states are often successfully employed to expand the set of attainable structures. Here we show that physical vapor deposition of liquid crystalline materials is a particularly promising route to highly organized nonequilibrium molecular solids.

Liquid crystalline (LC) states have often been utilized to produce molecular organization in solids.[1,6–8] For these systems, there are well-developed methods to control structure in the fluid state and molecular alignment can then be transferred into the solid state. Typically, thermal annealing and specially prepared substrates[9–13] are used to prepare large LC domains in equilibrium and subsequent cooling yields a solid, either crystalline or glassy, that inherits structural organization from the LC phase. In optoelectronics, for example, solid-state lasers with multiple lasing wavelengths in the same device are made by tuning the LC order (via composition) and then quenching into a glassy solid.[1] For organic electronics, both smectic (forming two dimensional molecular sheets) and discotic phases (forming one dimensional molecular columns), have been shown to promote anisotropic charge carrier mobility.[6,8,14] Enhancements in mobility have been observed in a number of highly aligned liquid-crystalline systems.[15,16] In one case, a discotic phase of a perylenediimide derivative[17] was reported to exhibit a 5 orders of magnitude enhancement in charge mobility along the direction of π-stacking upon alignment and similar performance improvements have been reported for other systems. [6] [18]

Physical vapor deposition, a technique used commercially in the fabrication of organic light emitting diodes (OLEDs),[19] can also be used to control molecular organization in solids. Glass films produced by this method are molecularly smooth, macroscopically homogenous and grain boundary-free, making them particularly useful in OLEDs and other multilayer device architectures.[20] Depending on deposition parameters, these glassy materials can be anisotropic, with tunable molecular orientation,[21,22] polar order,[23] and anisotropic molecular packing.[24,25] The structural order produced by vapor deposition, albeit small compared with that found in equilibrium LC systems, has been shown in one case to enhance charge carrier mobility by nearly an order of magnitude.[26] In contrast to the LC work described above, vapor deposition directly produces anisotropic order, without thermal annealing steps.

Can the ability of physical vapor deposition to prepare mildly anisotropic solids from non-LC systems be combined with the structure-forming tendency of LC molecules to directly produce ordered molecular solids? There is precedent for using physical vapor deposition to create thin films of LC systems[27–29]. Particularly relevant is the work of Echher et al., which compared the electrical performance of vapor-deposited films with spin-coated and annealed films of a columnar LC system.[29] They reported that the columnar order was higher for vapor-deposited films than spin-coated films, leading to improved charge transport characteristics similar to those obtained by thermal annealing. In another recent study, it was shown that controlling the substrate temperature ($T_{substrate}$) during physical vapor deposition produced a remarkable control over molecular orientation in glassy films of itraconazole, a rod-like molecule that can form isotropic, nematic and smectic phases;[30] the average tilt angle in the as-deposited solids varied from 27˚ to 76˚ relative to the film normal as a function of $T_{substrate}$. These films were found to be macroscopically homogeneous and molecularly smooth, making the process potentially useful in device fabrication. In contrast to the behavior of equilibrium LC systems, molecular orientation in vapor-deposited films of itraconazole did not depend upon the identity of the underlying substrate.

Here we use wide angle x-ray scattering (WAXS) and grazing incidence wide angle x-ray scattering (GIWAXS) to characterize the molecular packing of itraconazole glasses prepared by vapor deposition at various substrate temperatures. For $T_{substrate}$ near and below $T_g$, the glasses were found to have highly aligned smectic-like layering propagating through the film along the surface normal, independent of the total thickness of the film (in the range 100 – 1800 nm). Remarkably, glasses prepared by vapor deposition above $T_g$ exhibit much more organized smectic layering than could be produced by extended thermal annealing in the equilibrium smectic phase. The layer spacing could be controlled through $T_{substrate}$ and ranged from 30 Å to 25 Å. Combining this x-ray data with previous measurements of average molecular orientation,[30] we describe the microstructures of the various glasses produced. Our results are consistent with the mechanism advanced in ref [21], where preferred molecular orientation in a deposited glass is a combined result of  the high molecular mobility observed on free surfaces of glasses[31,32] and the tendency for molecules to self-assemble into anisotropic structures at the free surface of a liquid[21]. This mechanism and our experimental results indicate that molecular organization in the deposited glass is independent of the anchoring preferred at the substrate.

These results suggest that vapor deposition of LC systems can be a general route for the production of highly organized organic solids. Although our test system is a smectic LC, the proposed mechanism indicates that highly organized columnar and nematic phases should also be produced by vapor deposition, because these systems are also known to self-organize at the free surface.[33–35] Vapor deposition might be a particularly useful approach for columnar systems such as those currently being explored for use in organic electronics. In comparison to methods based upon thermal annealing, vapor deposition offers two key advantages: 1) as-deposited samples can be more ordered than those produced by thermal equilibration, and 2) multi-layer devices can be prepared without concern that thermal annealing of one layer will disrupt the packing of an underlying layer.

## Experimental Methods

**Sample preparation.** The structure of itraconazole is shown in Figure 1; the name "itraconazole" has been used in the literature to refer to any subset of the 8 stereoisomers of this structure.[36] The material utilized in our experiments was purchased from Sigma Aldrich (>98% purity) as a 1:1:1:1 racemic mixture of the 4 cis- stereoisomers[36] and was used as-received; in this paper, "itraconazole" always refers this mixture. The systematic name for this group of stereoisomers is cis-2-sec-butyl-4-[4-(4-{4-[2-(2,4-dichlorophenyl)-2-(1H-1,2,4-triazol-1-ylmethyl)-1,3-dioxolan-4-ylmethoxy]phenyl}-piperazin-1-yl)phenyl]-2,4-dihydro-1,2,4-triazol-3-one. Itraconazole is an isotropic liquid above 363 K. As the temperature is lowered, it exhibits a nematic phase down to 346 K and then exhibits a smectic phase which undergoes a transition to a smectic glass at $T_g$ = 330 K cooling at 1 K/min.[37] (See Supplementary Information Figure SI1 for DSC thermograph). Itraconazole is a good glass-former and can be maintained below its melting point (439 K) for extended periods of time without nucleation.

Glasses of itraconazole were prepared by physical vapor deposition onto silicon <100> substrates with a native oxide. The deposition rate was 0.2 ± 0.02 nm/s in a vacuum environment ($10^{-7}$ torr). The deposition rate was monitored using a quartz crystal microbalance. All the vapor-deposited glasses of itraconazole, for all $T_{substrate}$ values, were macroscopically homogeneous and pin-hole free.[30]

For the samples investigated by asymmetric wide-angle x-ray scattering (WAXS), a temperature gradient setup was used, as previously described,[22] to create a library of glasses prepared at different substrate temperatures. This is achieved by having two independently controlled temperature stages bridged by a rectangular substrate. The temperature differential between the two stages creates a temperature gradient along the silicon substrate that is maintained throughout the deposition process. The precise substrate temperature associated with various locations on the substrate was established by comparison with samples prepared at a single temperature; we find no discernable difference between the properties of a film prepared isothermally and a sample prepared at the same temperature on a temperature-gradient substrate.

After deposition, the higher temperature stage was cooled to room temperature at 1 K/min. The absolute error in substrate temperature is estimated to be less than 2 K. The samples used for WAXS measurements were 1.8 microns thick.

For the samples investigated by grazing-incidence wide-angle x-ray scattering (GIWAXS), itraconazole was deposited onto isothermal substrates. For $T_{substrate}$ above room temperature, the temperature stage was cooled to room temperature at 1 K/min after deposition. Both 100 nm and 300 nm thick samples were prepared for GIWAXS.

To compare the structure of a vapor deposited film with a thermally annealed film, a 1.8-micron thick vapor-deposited glass film was heated into the isotropic liquid, quenched to room temperature, and then annealed for 7 days at 337 K (where the equilibrium state is the smectic phase). This sample was then cooled to room temperature for characterization by WAXS.

**High-throughput WAXS measurements.** Asymmetric WAXS measurements were conducted on a Bruker D8 Discover diffraction instrument equipped with a Cu K-α x-ray source ($\lambda$= 1.54 Å), a beam spot size of 2 mm and a VANTEC500 area detector. A knife-edge made of Inconel 625 was installed to block air scatter. For the 2D WAXS patterns and the q scans in Figure 2, the incident angle was fixed at $\theta$= 2.5° for all measurements. For each measurement, the sample was exposed to x-rays for 240 s (with a 50 W power output of x-ray source). The exposures were done at various points along the length of the high-throughput temperature gradient sample, with every point corresponding to a unique $T_{substrate}$. A similar procedure was used for the $\chi$ scans in Figure 4 and the 2D patterns in Figure 6 except that the incident angle was adjusted for every pattern by matching the Bragg condition associated with the 2$^{nd}$ order reflection along $q_z$ to within 0.01 degrees. The 2$^{nd}$ order reflection was used rather than the primary peak at $q_z$~0.2 Å$^{-1}$ so as to avoid the influence of background scattering. For a list of angles and corresponding $T_{substrate}$, please see SI Table 1. The integration limits used to construct Figure 4 can be found in SI Table 2.

**GIWAXS measurements.** GIWAXS measurements were conducted on beamline 11-3 at the Stanford Synchrotron Radiation Lightsource (SSRL) with a wavelength of 0.973 Å. The incident angle was fixed to 0.14° such that the scattering occurred from the bulk of the film and samples were typically exposed for 200 seconds. The data was processed using the SSRL-developed WxDiff Integration Tool software. The diffraction patterns were corrected for polarization of the beam and $\chi$-corrected to obtain an accurate reciprocal space map, as described by Baker et al.[38]

# Results

**2D scattering patterns from vapor deposited glasses of itraconazole.** Figure 1a shows a subset of the x-ray scattering patterns obtained from glasses of itraconazole vapor-deposited at different $T_{substrate}$ values. The panels in Figure 1a were obtained from grazing incidence wide angle x-rays scattering (GIWAXS) from 300 nm thick films. The itraconazole glasses produced by vapor deposition exhibit remarkably diverse local packing structures as a function of $T_{substrate}$, as indicated in the schematics of Figure 1b. As described in more detail below, these range from highly ordered structures similar to smectic A packing, to a smectic C-like packing, and finally to a more disordered but highly anisotropic structure. Of these, only the smectic A structure is a known equilibrium state of itraconazole.

To aid in our analysis, a coordinate system associated with the scattering vector, q, is defined in $Å^{-1}$ for the 2D GIWAXS patterns, with $q_z$ corresponding to out-of-plane scattering plotted along the vertical axis and $q_{xy}$ corresponding to in-plane scattering plotted along the horizontal axis.[38] The scattering vector is defined by $q = 4\pi \sin(\theta)/\lambda$, where $\theta$ is half the scattering angle and $\lambda$ is the x-ray wavelength. $q_x$ and $q_y$ are equivalent since rotating the sample in the xy-plane led to no significant change in the scattering pattern. The in-plane isotropy of these glasses has also been confirmed using optical microscopy and atomic force microscopy, neither of which show domain structure on the observable length scales.[30]

Glasses deposited at $T_{substrate} = 335$ K = $T_g + 5$ K (top row of Figure 1) show the scattering features expected for a highly aligned smectic A liquid crystal. Sharp diffraction peaks are seen in the grazing incidence wide-angle x-ray scattering (GIWAXS) pattern at around $q_z{\sim}0.2$ $Å^{-1}$, with higher order peaks appearing at $q_z{\sim}0.4$ $Å^{-1}$, and faintly at $q_z{\sim}0.6$ $Å^{-1}$. (There is a weak feature at $q_z{\sim}0.3$ $Å^{-1}$ due to sample reflectivity.) This sequence of peaks indicates a periodic structure with long-range order and a large correlation length out of the plane of the substrate, with a periodicity of about d = 30 Å ($d = 2\pi/q$). This length-scale is similar to that observed in the equilibrium smectic liquid structure and to the molecular length (33 Å by DFT[37] and 30 Å from single crystal x-ray studies[39]). It is consistent with end-to-end stacking of the molecules in a smectic microstructure. The GIWAXS measurements also show a large in-plane scattering peak at $q_{xy}{\sim}1.4$ $Å^{-1}$. This indicates the lateral (side-by-side) packing expected for rod-like molecules in smectic layers, with a nearest-neighbor spacing d ~ 4.5 Å. The GIWAXS results (and the asymmetric WAXS results discussed below) are consistent with the view that for $T_{substrate}$>$T_g$, the films are in the equilibrium smectic phase during deposition. Upon cooling to room temperature, this smectic structure gets trapped in the glassy state, i.e., the glass structure is inherited from the equilibrium smectic liquid. The most remarkable feature of glasses deposited near 335 K is the strong alignment of the smectic layers parallel with the substrate, as we discuss below.

Glasses deposited at $T_{substrate}$ = 330 K = $T_g$ (second row of Figure 1) are similar to the those formed above $T_g$. The out-of-plane peaks observed in the GIWAXS pattern are in the same position along $q_z$. However, these peaks are broader in $q_{xy}$ than in the glass described above. Qualitatively, this kind of peak broadening is consistent with some small lateral disorder, such as modulations in the layering.[40,41]

Glasses deposited at $T_{substrate}$ = 315 K = $T_g$ – 15 K (third row in Figure 1) show qualitatively similar out-of-plane diffraction peaks as described above but they are shifted to higher values along $q_z$, indicating smaller layer spacings in these glasses (due to an increase in the average tilt of the molecules). The out-of-plane peaks also broaden azimuthally (along $\chi$, see Figure 4 inset for schematic definition of $\chi$), indicative of a textured structure within the film, i.e., a distribution of molecular layer orientations. The azimuthal peak broadening described here is likely convolved with lateral (along $q_{xy}$) broadening as we discuss below. The in-plane diffraction peak seen in the GIWAXS at q~1.4 $A^{-1}$ along $q_{xy}$ also broadens azimuthally, consistent with a slightly more orientationally disordered (textured) in-plane structure than for the glasses prepared at $T_{substrate}$ ~ $T_g$. This structure is similar to some equilibrium smectic C systems.[42]

A much different diffraction pattern is seen for itraconazole glasses formed with $T_{substrate}$ = 260 K (fourth row in Figure 1) as compared to the higher $T_{substrate}$ films. The out-of-plane peaks at $q_z$~0.2, 0.4 and 0.6 Å$^{-1}$ are no longer present and a broad peak is seen at $q_z$~1.4 Å$^{-1}$, indicating that the molecules in the film are predominantly lying in the plane of the substrate. Broad low-intensity peaks are observed in the GIWAXS at low $q_{xy}$ values, consistent with a nematic or weakly smectic-like packing in-plane. (A line cut along $q_{xy}$ is provided in Supplementary Information Figure SI2). Based on the x-ray and optical evidence for in-plane isotropy for all the PVD glasses, any in-plane smectic layer propagation must be randomly distributed in-plane.

**The thermodynamic state of the as-deposited films**. The films of itraconazole prepared by vapor deposition are out-of-equilibrium (glassy) solids at room temperature (where all the x-ray measurements are performed). They are non-crystalline as evidenced by the observation that not all the x-ray diffraction peaks are sharp. The as-deposited glasses are solids that can be temperature-cycled below $T_g$ with no change in structure. The films prepared by depositing at and above $T_g$, upon cooling to room temperature, are similar to glassy liquid crystals obtained by cooling the equilibrium smectic liquid,[43] as studied previously.[1,44] (Although not all glassy smectic liquid crystals will exhibit the high level of alignment observed in these samples as we discuss below.) The microstructures of the films prepared by depositing at $T_{substrate}<T_g$, however, are quite different than a glass prepared by cooling the equilibrium smectic liquid. When these samples are heated to about $T_g$ + 15 K the as-deposited structure is lost, and the equilibrium smectic liquid state is formed. Upon cooling to room temperature, a glassy smectic is formed, with its structure inherited from the equilibrium smectic state. The observation that the as-deposited structure is irreversibly lost upon thermally cycling just above $T_g$ further indicates the films are not crystalline as itraconazole does not melt until a much higher temperature (439 K).

**Tuning smectic-like layer spacing by choice of $T_{substrate}$.** As introduced in Figure 1, the choice of $T_{substrate}$ in a range just below $T_g$ allows the preparation of itraconzole glasses in which the spacing of the smectic-like layered structures is controlled. This is investigated more closely in Figure 2 using the asymmetric WAXS data. Representative 2D patterns are shown at the top of the figure. In these panels, one can observe the first and second order reflections corresponding to the smectic layers (and sometimes also the third order reflection). (The WAXS patterns also show a faint reflectivity peak at $q_z$~0.3 $Å^{-1}$, and background intensity below 0.2 $Å^{-1}$ caused by the low incident angle geometry of these measurements.)

The main panel of Figure 2 illustrates the spacing of the smectic-like layers by showing a line cut from the WAXS data along q for various glasses of itraconazole deposited between 337 K and 300 K. These data were obtained by integrating over an azimuthal angle approximately the width of the WAXS patterns at the top of the figure. For comparison, data is also presented for as a vapor-deposited sample that was heated into the isotropic liquid state and then annealed at 337 K for one week. All the x-ray scattering measurements shown in Figure 2 were conducted at room temperature.

The diffraction peaks shown in Figure 2 for the samples vapor-deposited at $T_{substrate} \geq 307$ K indicate smectic-like layering the samples. Each of these samples shows the fundamental peak (~0.23 $Å^{-1}$) and higher order reflections (~0.45 $Å^{-1}$ and sometimes 0.7 $Å^{-1}$). These peaks shift to higher q values and broaden systematically as $T_{substrate}$ is reduced. Higher q values indicate smaller layer spacing, while the broadening is a sign of shorter coherence lengths and more disorder in the layered structure.

The spacing associated with the smectic-like layers is shown as a function of the substrate temperature during deposition in Figure 3, with the glass transition temperature marked with a vertical black line at 330 K. Above $T_g$, the layer spacing remains nearly constant, consistent with the view that the film being deposited is in the equilibrium smectic phase before being cooled into a glassy solid. For samples directly deposited into a glassy state below $T_g$, however, the layer spacing is systematically and monotonically reduced by up to 16%, from ~30 to ~25 Å. A broadening of the diffraction peaks along q is also observed at lower substrate temperature during deposition, suggestive of less long-range order in films and perhaps less layer spacing uniformity. This is quantified by the peak width along q and is shown in supplementary information Figure SI3.

**Alignment of smectic-like layers in vapor-deposited glasses of itraconazole**. Figure 4 shows azimuthal line cuts across the second order diffraction peak at q~0.4 $Å^{-1}$ for vapor-deposited and thermally annealed glasses of itraconazole, and allows a quantitative comparison of the degree of alignment of the smectic-like layers in the different samples. The scattering intensity was collected at the Bragg condition for the peaks and is plotted against $\chi$, the azimuthal angle (defined such that $\chi = 0^\circ$ is along $q_z$ and $\chi = 90^\circ$ is along $q_{xy}$). Higher intensity and sharper peaks correspond to well-ordered structures with large coherence lengths and a narrow distribution of directors. Figures 4b shows the same data as Figure 4a but with an expanded vertical axis.

The most striking feature of Figure 4 is that several vapor-deposited itraconazole glasses show more highly aligned smectic layers than the thermally annealed sample, indicating that these vapor deposited films are more highly ordered. In particular, the scattering intensity for the sample directly deposited at 337 K is more than one order of magnitude larger than for the thermally annealed sample. The annealed sample was obtained by heating a highly ordered vapor-deposited glass into the isotropic liquid, quenching rapidly, annealing for 7 days at 337 K (where the equilibrium phase is smectic), and finally cooling at 1 K/min to ambient conditions. In contrast, the highly aligned sample vapor-deposited at 337 K required only 2.5 hours to prepare at the 2 Å/s deposition rate. We further discuss the comparison between vapor-deposited and annealed samples below.

For the vapor-deposited glasses shown in Figure 4, the degree of smectic order is much lower when the substrate temperature during deposition is lower. Figure 4b shows, in addition, that the peaks arising from glasses deposited at $T_{substrate} < T_g$ broaden significantly. These features indicate that depositions at lower $T_{substrate}$ lead to layered structures with smaller coherence lengths. Qualitatively, the peak broadening observed in the second order peak may arise from at least two types of imperfections. 1) Peak broadening along the azimuthal angle ($\chi$) indicates that the layers are not entirely parallel to the substrate but rather exhibit a textured structure. 2) Lateral peak broadening (along $q_{xy}$) is indicative of in-plane defects. We have not attempted to distinguish between these two explanations and we note that Figure 4 would appear very similar if the linecut had been taken along $q_{xy}$ rather than along $\chi$.

**Probing lateral molecular packing.** The GIWAXS scattering data near q =1.4 Å$^{-1}$, as discussed in the description of Figure 1, contains information about nearest-neighbor packing in vapor-deposited itraconazole glasses. For example, the very strong feature along $q_{xy}$ in Figure 1a is expected from the scattering of closely-packed, nearly vertical rod-like molecules. Figure 5 plots an orientation order parameter $S_{N\text{-}N}$ based upon this feature in red; $S_{N\text{-}N}$ was measured from the GIWAXS patterns as described previously.[25] A value of $S_{N\text{-}N}$ = +1 would indicate molecules packing parallel to each other in the plane of the substrate, while $S_{N\text{-}N}$ = -0.5 would indicate molecules parallel to each other and normal to the substrate. For films deposited at $T_{substrate} \geq T_g$, there is a strong tendency for the molecules to pack normal to the substrate, consistent with smectic-like layers parallel to the substrate. For films deposited a little below $T_g$, the observed packing becomes more isotropic. For glasses grown at $T_{substrate} \ll T_g$, $S_{N\text{-}N}$ is positive, indicating a tendency for the molecules to pack with their long axes laying in the plane of the substrate and parallel to each other. Although the maximum value of $S_{N\text{-}N}$ = +0.12 is quite small, it is twice the value reported for glasses of TPD vapor-deposited at low substrate temperatures.[25]

Figure 5 also plots an FTIR-based orientation order parameter $S_{FTIR}$ as described in detail elsewhere.[30,37] Based on IR dichroism, $S_{FTIR}$ is a measure of the average molecular orientation of an axis very nearly parallel to the long axis of itraconazole molecules in the glass. $S_{FTIR}$ = +1 would indicate molecules standing vertically while $S_{FTIR}$ = -0.5 would indicate that all the molecules are lying in the plane of the substrate. Figure 5 shows a strong anti-correlation between $S_{N\text{-}N}$ and $S_{FTIR}$. For $T_{substrate} \geq T_g$, $S_{FTIR}$ indicates that the molecules have a strong tendency to stand upright, consistent with the negative $S_{N\text{-}N}$ value and a smectic structure with planes parallel to the substrate. For $T_{substrate}$ near 260 K, $S_{FTIR}$ indicates a tendency for the molecules to lay nearly flat in the plane, in agreement with the positive $S_{N-N}$ value. An interesting point of comparison is the $S_{FTIR}$ value measured by Tarnacka and coworkers[37] for itraconazole in a homeotropically aligned smectic phase (plotted as the purple star in Figure 5). The good agreement of this value with $S_{FTIR}$ for the vapor-deposited samples is consistent with direct deposition into a highly aligned smectic structure.

**Comparison of thermally annealed and vapor-deposited itraconazole films.** Figure 6 shows the intriguing result that the itraconazole sample vapor-deposited at 337 K has more highly organized smectic layers than a sample thermally annealed at 337 K for one week. Initially, we were surprised that a non-equilibrium assembly process would be more effective in producing smectic alignment than extensive thermal equilibration. Figure 6 shows a comparison of WAXS data for these two samples that allows further insight into this issue. The scattering at $q \sim 1.4$ Å$^{-1}$ along a wide range of $\chi$ in the thermally annealed sample would not be present in a smectic phase with planes propagating exclusively along the surface normal and this feature is indeed absent from the vapor-deposited sample. For the thermally annealed sample, the slightly larger scattering at $q=1.4$ Å$^{-1}$ around $\chi=0^{\circ}$ is consistent with a smectic population with planar anchoring (i.e., the molecular long axis lies along the interface). The differences in the orientation of smectic packing are schematically illustrated in Figure 6. We expect that the silicon substrate with the native oxide promotes planar anchoring[45] and, for the thermally annealed sample, this competes with the homeotropic anchoring (i.e., the molecular long axis lies nearly normal to the interface) preferred at the free surface. With this interpretation in mind, we can understand that the peak at q ~ 0.4 $Å^{-1}$ along $\chi$=0° is weaker in the thermally annealed sample because it represents only the fraction of the smectic layers that propagate only the surface normal.

**Structure.** We present schematic structures for vapor-deposited itraconazole glasses in Figure 7 for three representative substrate temperatures, based upon the x-ray measurements presented here and results previously reported in ref [30]. The spacing of the smectic-like structures shown in the top two panels comes from the values reported in Figure 3. An average molecular orientation at each substrate temperature can be determined from $S_{FTIR}$, as plotted in figure 5:[30]

$$\theta_{avg} = \arccos[(2/3 S_{FTIR} + 1/3)^{1/2}]$$

Here $\theta_{avg}$ is measured from the substrate normal. It is important to note that, for the smectic-like structures prepared by vapor deposition, a collection of the structures shown in Figure 7 with a narrow distribution of molecular layer orientations is likely present across the film; the wider the distribution of layer orientations, the broader the azimuthal spread in the out-of-plane x-ray diffraction peaks. While the layers on average always propagate perpendicular to the substrate leading to out-of-plane diffraction peaks, structures like those shown in Figures 7 a and b likely have random in-plane orientation, leading to in-plane isotropy on the length-scales probed here and in microscopy[30] studies. At the lowest values of $T_{substrate}$, layers are not formed, and instead, the molecules lie nearly flat in the plane of the substrate. To the best of our knowledge, the tunable layer spacing and highly variable molecular orientation exhibited by vapor-deposited itraconazole glasses is unprecedented.

The microstructures shown in Figure 7 persist throughout the thickness of the vapor-deposited itraconazole glasses. The peak positions in the x-ray scattering patterns are consistent for films ranging from around 100 nm to 1.8 microns, indicating that the spacing of the smectic-like layers is independent of thickness. The orientation order parameter for nearest-neighbor packing $S_{N\text{-}N}$ is the same for 100 nm and 300 nm films. Furthermore, ref. [30] shows that the optical birefringence of the vapor-deposited glasses is the same for films between 180 nm and 650 nm, while $S_{FTIR}$ is independent in the range of 330 nm to 1800 nm.

## Discussion

In this section, we address a number of questions about the mechanism by which vapor deposition produces ordered glasses of itraconazole and, in particular, how to understand the strong influence of substrate temperature on the type of order trapped in the glass (Figure 7). We also discuss how non-equilibrium deposition can lead to more highly ordered solids than thermal equilibration (Figure 6). Finally, we describe how the unique structures that can be prepared by vapor deposition might be utilized.

**Mechanism of layer formation in itraconazole glasses.** We begin by briefly reviewing what is known about vapor deposition of molecules that do not form liquid-crystals. Vapor deposition onto substrates near 0.85 $T_g$ produces high density, low enthalpy glasses of many organic molecules.[22,46] High mobility near the free surface during deposition allows nearly complete equilibration of just-deposited molecules into tightly packed configurations; further deposition traps these molecules into place. Clearly surface equilibration will also occur for deposition with $T_{substrate} = T_g$, but under these conditions there is no thermodynamic driving force to form a high density material. At low substrate temperatures ($T_{substrate} << 0.85T_g$), surface mobility will be insufficient to reach high density states even though they are thermodynamically favored.

The above mechanism for the formation of well-packed glasses is sufficient to explain the structures shown in Figure 7 if we add one new element, as discussed previously.[30] At the equilibrium free-surface of smectic and nematic forming liquids, molecules favor a smectic-like structure, homeotropically anchored at the vacuum interface.[33,47,48] With this mechanism, deposition onto substrates at $T_g$ will lead to a smectic structure with a single director normal to the substrate (Figure 7a); every new layer of molecules wants to form a smectic structure anchored to the free surface and each layer can fit coherently on top of the previous one. Deposition at $T_g$ + 7 K leads to structures with even higher perfection because the bulk material further equilibrates during the deposition process. At $T_g$ + 7 K, structural rearrangements in the bulk take place on the time scale of 1 second while the deposition requires 500-10000 seconds, depending upon film thickness; for comparison, molecular motions are nearly 100 times slower at $T_g$.[37,49] When deposition occurs onto substrates just below $T_g$ (middle panel in Figure 7), molecules near the free surface of the film still have a driving force to achieve the equilibrium smectic layer but not quite enough time to do so. An incompletely organized smectic layer with the “wrong” layer spacing gets trapped into place by further deposition since at these temperatures reorganization in the bulk material is negligible. Since mobility at the surface depends strongly on temperature, the system more closely attains its equilibrium structure at higher values of $T_{substrate}$, as shown in Figure 3.

At $T_{substrate}$ near 0.8 $T_g$, surface mobility is so low that just-deposited molecules make no progress in their attempt to create homeotropically-anchored smectic layers. In this regime, molecules tend to lie parallel to the substrate; we expect that this minimizes the energy for an incoming molecule landing on a relatively flat surface.[21] Interestingly, the molecules in these glasses have a slight tendency to form smectic-like or nematic-like structures with directors in the plane of the substrate. This can be seen in Figure 1a in the 2D GIWAXS scattering pattern of $T_{substrate}$ = 260 K. Along $q_{xy}$, small diffuse peaks appear at around 0.24, 0.48 and 0.72 Å$^{-1}$, indicating a periodic structure in the plane of the substrate; see supplementary figure SI2. This feature is isotropic in the plane, indicating the presence of weakly-organized domains of randomly oriented in-plane layers.

The above mechanism allows an understanding of the surprising result that vapor-deposited glasses can be more ordered than thermally equilibrated samples, as shown in Figure 6. A key feature of the proposed mechanism is that molecular organization in the as-deposited glass is being completely controlled by the free surface; $T_{substrate}$ determines the mobility near the free surface and this determines the extent to which the equilibrium surface structure is attained. In contrast, during thermal equilibration in the smectic regime, molecular organization of a thin film depends upon the preferred structure at both the free surface and the substrate. For the substrates used in these experiments, we expect that planar anchoring is preferred. Thus the bulk sample equilibrates as best it can with incompatible boundary conditions at the two interfaces and achieves a somewhat disordered compromise as represented by the schematic in Figure 6.

The proposal that smectic alignment for vapor-deposited itraconazole is completely controlled by the free surface is supported by preliminary deposition experiments on various substrates. We have observed that deposition onto silicon substrates treated to promote homoetropic anchoring has no effect on the structure of the vapor-deposited glasses. In addition, deposition of itraconazole at 325 K onto an already-deposited layer prepared at 270 K results in strong smectic layering with the director along the surface normal, even though molecules in the underlying layer are nearly planar. This conclusion expands upon birefringence results reported in reference [30] indicating that the structures of the vapor-deposited itraconazole glasses prepared here do not depend on the identity of the underlying substrate. Our assumption of high mobility at the surface of itraconazole glasses is consistent with previous work on organic glass formers indicating up to 8 orders of magnitude enhancement of surface diffusion relative to the bulk;[31,32] this assumption could be tested experimentally.

While not investigated here, it is likely that the vapor-deposited glasses of itraconazole exhibit polar order, as observed previously for deposition of polar molecules.[23,50] The mechanism described above implies that glasses deposited at different substrate temperatures could exhibit different levels of polar order and this warrants further investigation.

In comparison with previous literature on vapor-deposited LC systems, the current results provide evidence of both more highly ordered samples and tunable smectic spacing. In earlier studies of vapor-deposited thiophene, the microstructures achieved in the prepared films were similar to those of the known equilibrium smectic state; since only out-of-plane x-ray scattering was reported, it is not possible to characterize the extent to which the smectic layers were oriented along the surface normal. Variable layer spacing at equilibrium has been reported for some systems that exhibit a regular smectic A-to-smectic C transition.[42] In this case, the equilibrium layer spacing is reduced with temperature over a range of about 10%. While solids with variable smectic spacing might be formed by quenching these materials, we are unaware of such reports.

**Potential applications.** As discussed in the introduction, organized solids prepared via LC states are being widely explored for use in organic electronics and optoelectronics. Glassy liquid crystals have already been used as thin film transistors, solid-state lasers, optical memory, phosphorescent organic light emitting diodes (PhOLEDs) and, optical devices such as polarizers and notch filters.[1] For these applications, active layers are prepared by thermal annealing to attain an equilibrium LC state followed by cooling to form the liquid crystal glass.

Relative to processes that rely on thermal annealing, vapor-deposition expands the range of molecular packing arrangements and orientations available for organic electronic and optoelectronic applications. Deposition allows preparation of non-equilibrium microstructures such as smectic layers with tunable layer spacings, as shown in Figure 3. In addition, deposition can directly prepare highly oriented smectic structures that are functionally monodomains, even on substrates which would not support such high levels of organization via thermal annealing, as shown in Figure 6. Previous work has established that vapor-deposited glasses of itraconazole are macroscopically homogeneous and molecularly smooth, as required for many applications.[30] Vapor deposition is already used on a massive scale to produce organic light emitting diodes for cellphone displays and this illustrates the compatibility of deposition with high throughput manufacturing. In contrast, the preparation of multilayer devices by thermal annealing may be time-consuming and could be complicated both by the influence of underlying layers on molecular anchoring and the possibility that annealing one layer might degrade the structure of an underlying layer.

## Concluding remarks

Glasses of itraconazole, a model LC system, were formed by physical vapor deposition. Glasses with a range of interesting microstructures can be prepared by controlling the substrate temperature during the deposition process. Deposition onto substrates above $T_g$ yielded highly aligned smectic layers with the director normal to the substrate; these samples were better aligned than a thermally annealed sample. Deposition just below $T_g$ allowed the spacing of the smectic-like layers to be controlled over a range of 16%. At the lowest substrate temperatures investigated, molecules preferentially lay nearly flat in the plane of the substrate. These microstructures are formed throughout the thickness of the film and are stable at temperatures below $T_g$.

The mechanism proposed here to explain the observed microstructures suggests that similarly interesting structures will be formed by vapor deposition of other LC systems such as those used in organic electronics. Key elements of the mechanism are the smectic organization preferred by molecules at the free surface of the glass and enhanced molecular mobility near the free surface. An important consequence of this mechanism is that the anchoring preferred at the substrate is irrelevant. Vapor deposition is thus a tool for the preparation of structures in LC glasses that cannot be obtained by thermal annealing. An important next step is to extend this work to other types of liquid crystals in order explore what novel structures can be attained in these systems through the control of substrate temperature. Columnar systems are a particularly interesting target as they have often been utilized for organic electronics because of the efficient charge transport that is possible along the columns.

Acknowledgements – WAXS experiments were supported by the National Science Foundation (DMR-1234320). GIWAXS experiments were supported by NSF through the University of Wisconsin Materials Research Science and Engineering Center (DMR-1121288), which also provided instrumentation support. K.O. and M.L.C. were partially supported by NSF DMR award 1410438. Use of the Stanford Synchrotron Radiation Lightsource, SLAC National Accelerator Laboratory, is supported by the U.S. Department of Energy, Office of Science, Office of Basic Energy Sciences under Contract No. DE-AC02-76SF00515.

# Figures

Figure 1.

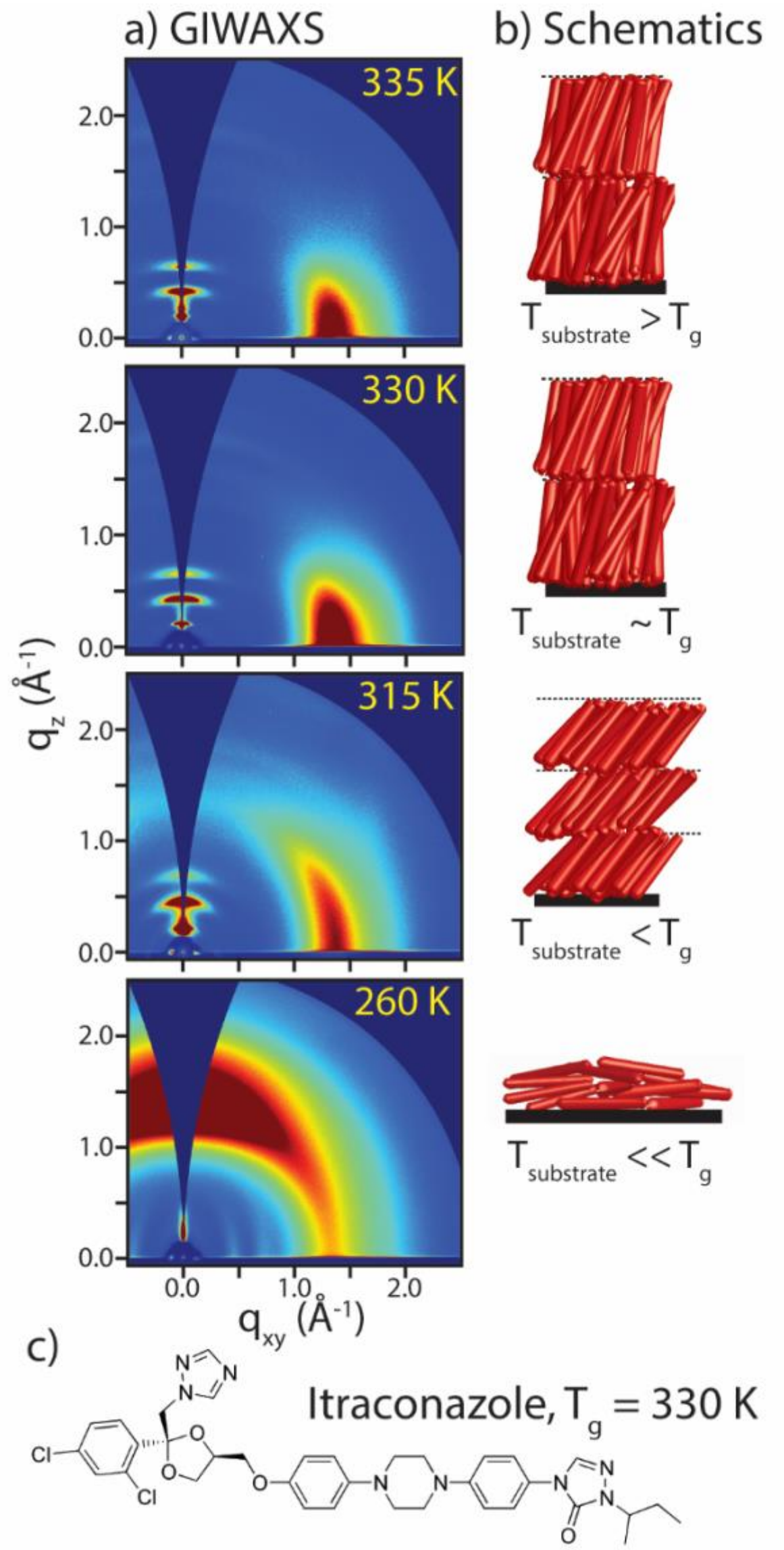

Figure 1. X-ray scattering patterns for glasses of itraconazole prepared by vapor deposition onto substrates with temperatures indicated. a) 2D GIWAXS scattering from 300 nm films. b) Schematics of proposed microstructures, with each red cylinder representing an itraconazole molecule, showing smectic-like layering for glasses prepared at high substrate temperatures. c) Molecular structure of itraconazole.

Figure 2.

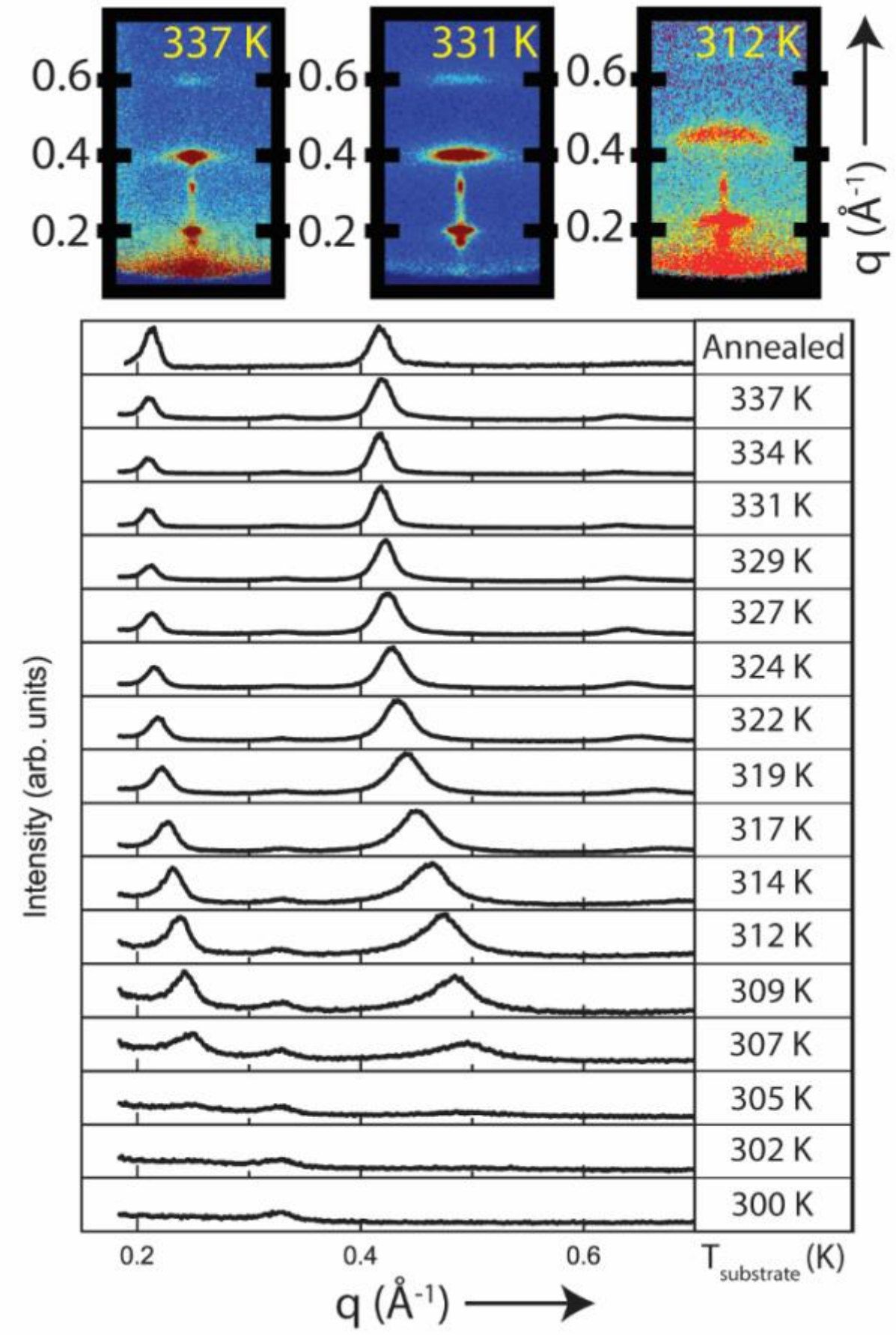

Figure 2. WAXS scattering intensities along vector q, as defined in top panel, for an annealed glass and various vapor-deposited glasses of itraconazole. Top: Representative 2D asymmetric WAXS patterns from itraconazole glasses deposited at $T_{substrate}$ indicated inset. The width of the pattern is approximately the integration area used to obtain the 1D diffractograms shown below. Bottom: Scattering intensity along q from WAXS for glasses of itraconazole deposited at the indicated substrate temperatures ($T_{substrate}$). The first panel shows data for a thermally annealed film. The shift of the scattering peaks to higher q at lower deposition temperatures indicates a decrease in the thickness of the smectic-like layers. All curves for vapor-deposited glasses have been normalized to the data presented for $T_{substrate} = 337$ K.

Figure 3.

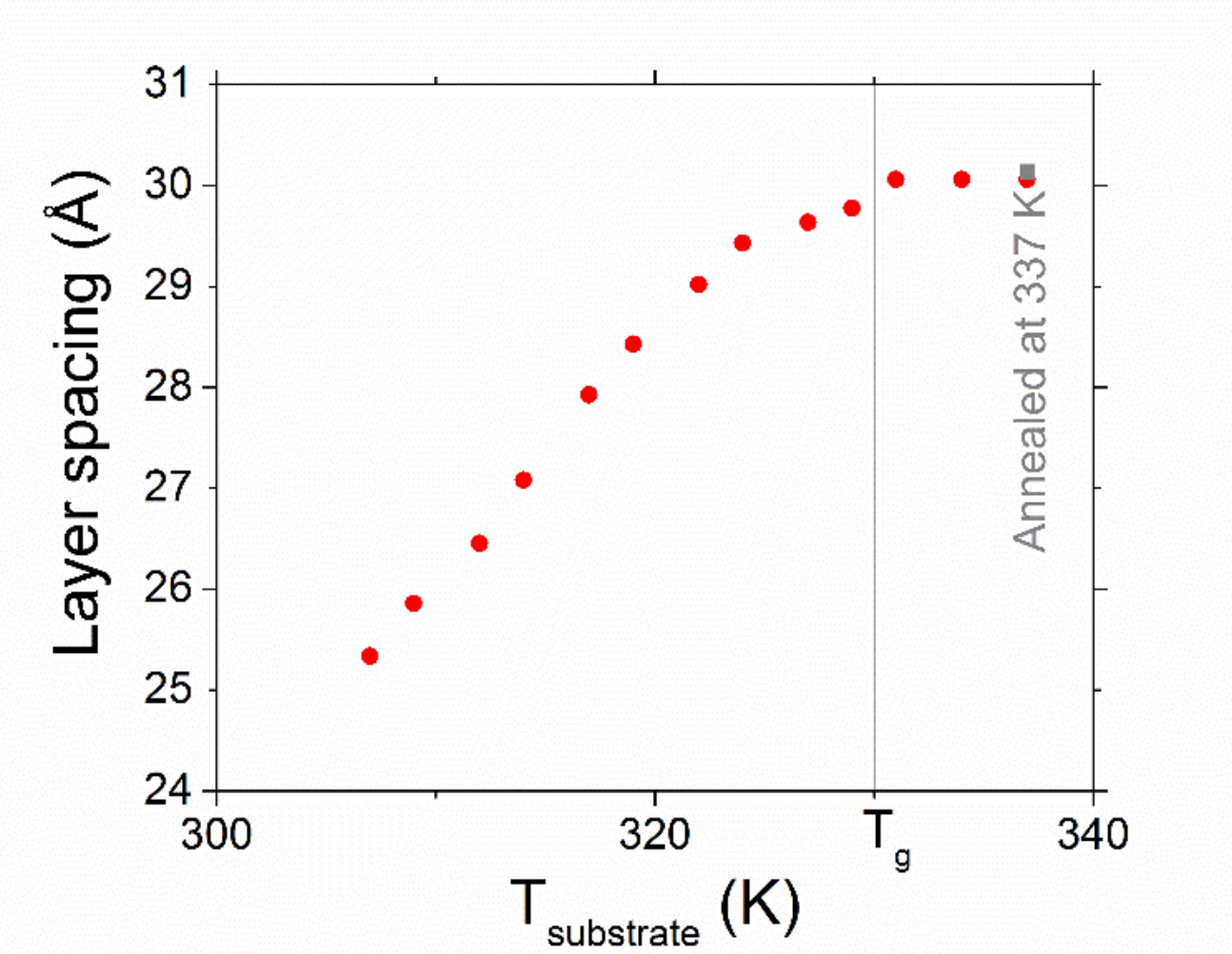

Figure 3. Thickness of smectic-like layers for vapor-deposited itraconazole glasses, as a function of substrate temperature during deposition ($T_{substrate}$), as derived from the position of scattering peaks along $q_z$. $T_{substrate} = T_g$ is demarcated with a vertical line. The layer spacing for a sample thermally annealed at 337 K is indicated by the gray square.

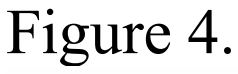

Figure 4.

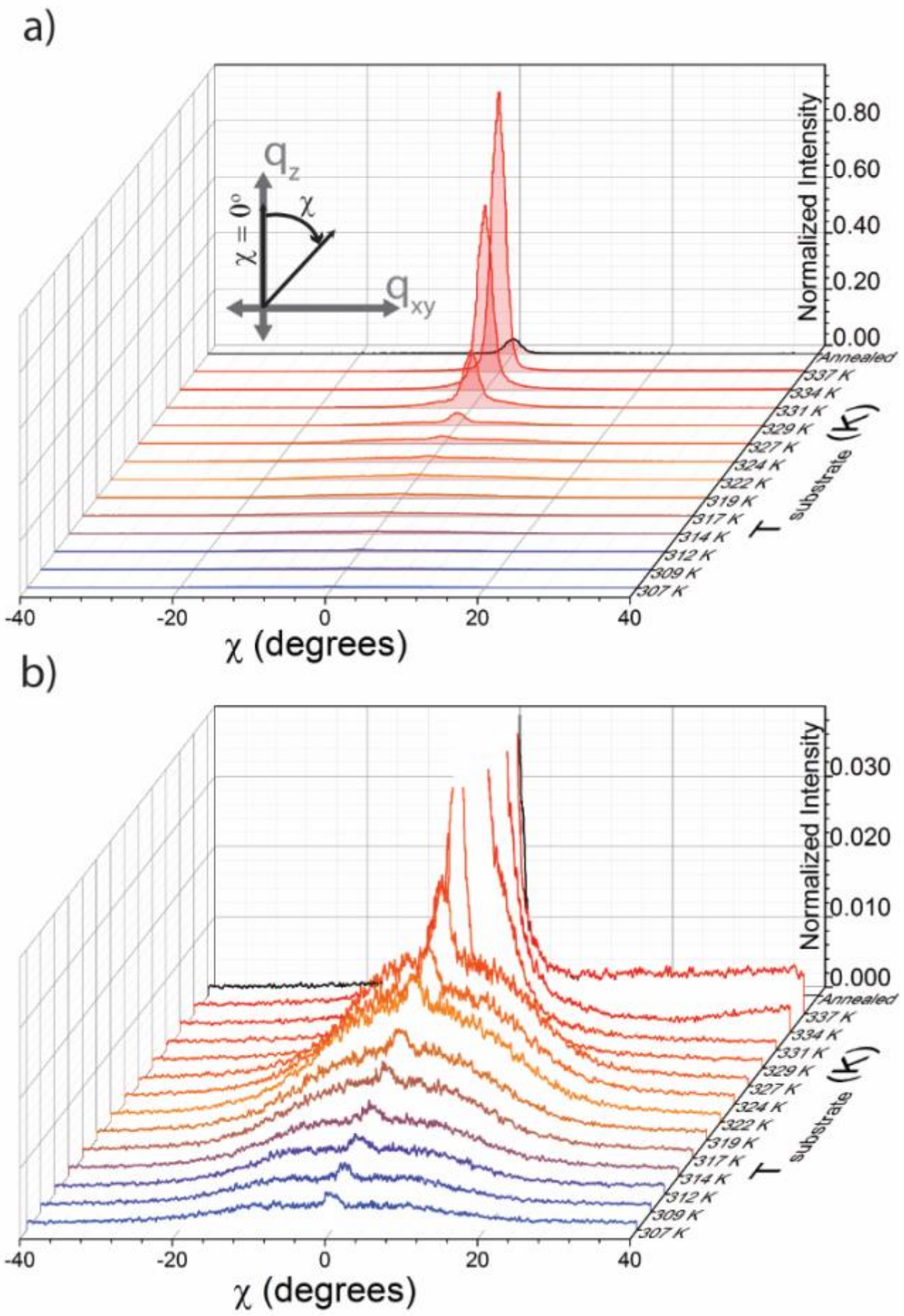

Figure 4. WAXS scattering data showing the 2nd order out-of-plane diffraction peak ($q_z$~0.4 Å$^{-1}$) along the azimuthal angle χ for itraconazole glasses deposited at various substrate temperatures ($T_{substrate}$), with χ defined in the inset. The intensity and width of this scattering feature characterize the quality of smectic organization and alignment. Multiple line cuts are shown offset in the z-axis, corresponding to different values of $T_{substrate}$. Panel b is a vertically expanded version of panel a. The black curve shows data for a sample annealed at 337 K. For both panels, intensities are normalized to the peak intensity observed for the film deposited at $T_{substrate}$ = 337 K.

Figure 5.

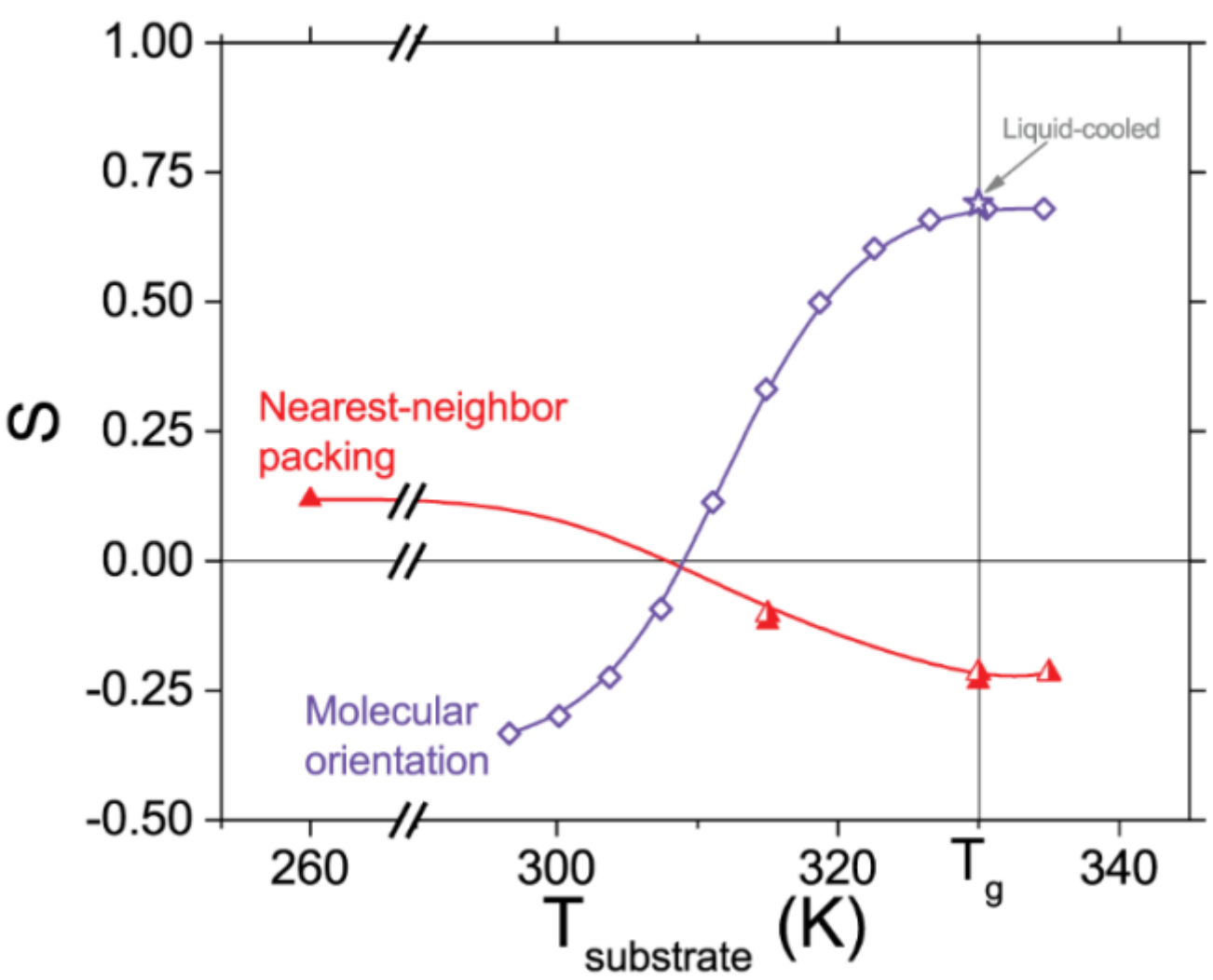

Figure 5. Orientation order parameters for nearest-neighbor packing ($S_{N\text{-}N}$) obtained from the GIWAXS data and for molecular orientation obtained from FTIR ($S_{FTIR}$), as a function of substrate temperature during deposition ($T_{substrate}$). $S_{N\text{-}N}$ quantifies the anisotropy in lateral nearest-neighbor molecular packing, varying from face-on at S = +1 to edge-on at S = -0.5; full triangles are 300 nm films and half-filled triangles are 120 nm films. $S_{FTIR}$ is a measure of the average orientation of the long molecular axis, varying from vertical at S = 1 to horizontal at S = -0.5; open diamonds are vapor-deposited glasses[30] and the purple star is for the equilibrium liquid[37]). The solid lines are guides to the eye.

Figure 6.

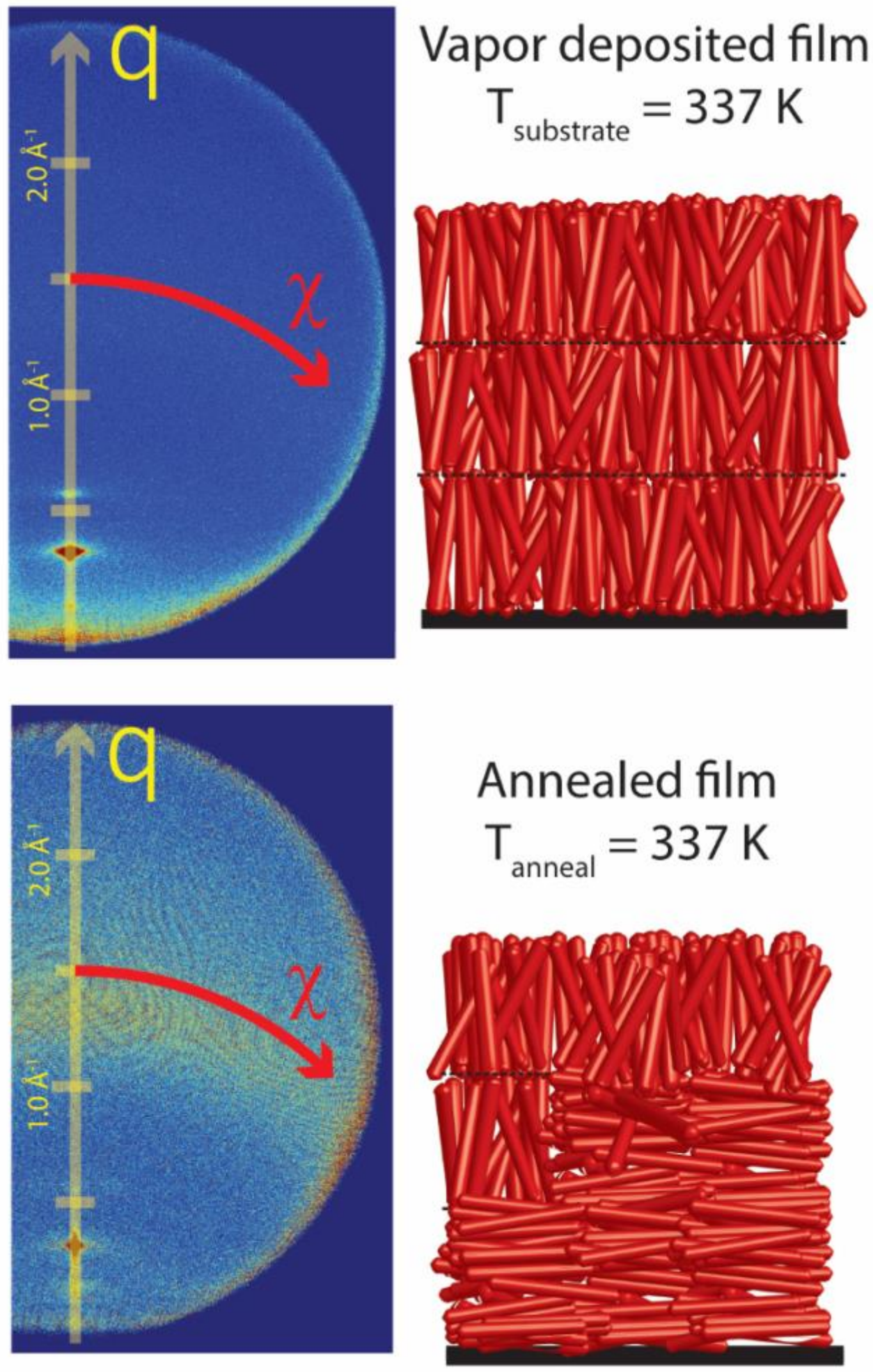

Figure 6. WAXS scattering patterns and possible structures for vapor-deposited and annealed films of itraconazole (1.8 microns thick). Top panel: 2D scattering pattern of film vapor-deposited at $T_{substrate}$ = 337 K with schematic structure. Bottom panel: 2D scattering pattern of film annealed for one week at $T_{anneal}$ = 337 K with schematic structure. The presence of the broad peak at q~1.4 $Å^{-1}$ in the annealed film indicates some population of molecules with planar alignment. In contrast, the vapor-deposited film shows no evidence of planar alignment.

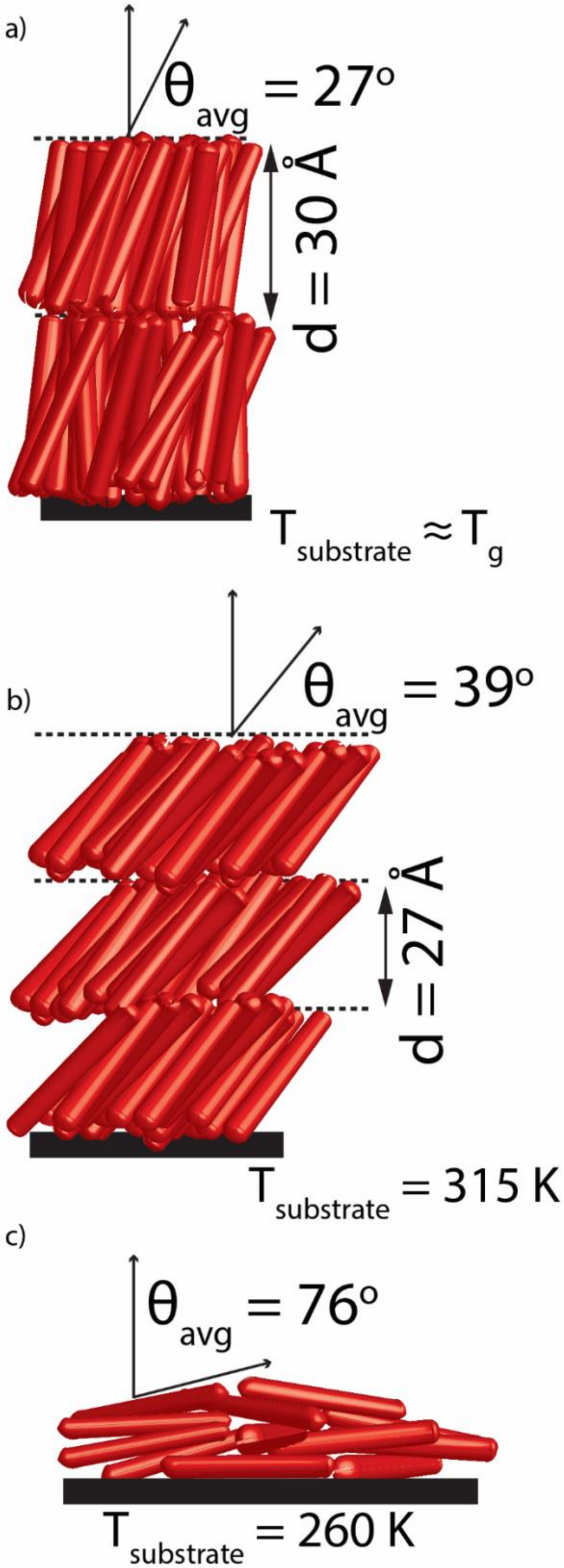

Figure 7. Schematics of a representative subset of microstructures for itraconazole glasses deposited at different substrate temperatures: a) $T_{substrate} \sim T_g$; b) $T_{substrate}$ = 315 K; c) $T_{substrate}$ = 260 K. While the structure in panel a is very similar to a highly aligned equilibrium smectic liquid, the structures shown in panels b and c have no equilibrium analogues for itraconazole. The values of the average tilt angle and the average layer spacing are derived from FTIR and WAXS as described in the text.

# Tunable smectic-like packing in vapor-deposited glasses of a liquid crystal

## Supplementary Information

*Ankit Gujral, Jaritza Gómez, Jing Jiang, Chengbin Huang, Kathryn A. O'Hara, Michael F. Toney, Michael L. Chabinyc, Lian Yu, M.D. Ediger*

**SI Figure 1. DSC thermograph of itraconazole obtained upon heating from glass**

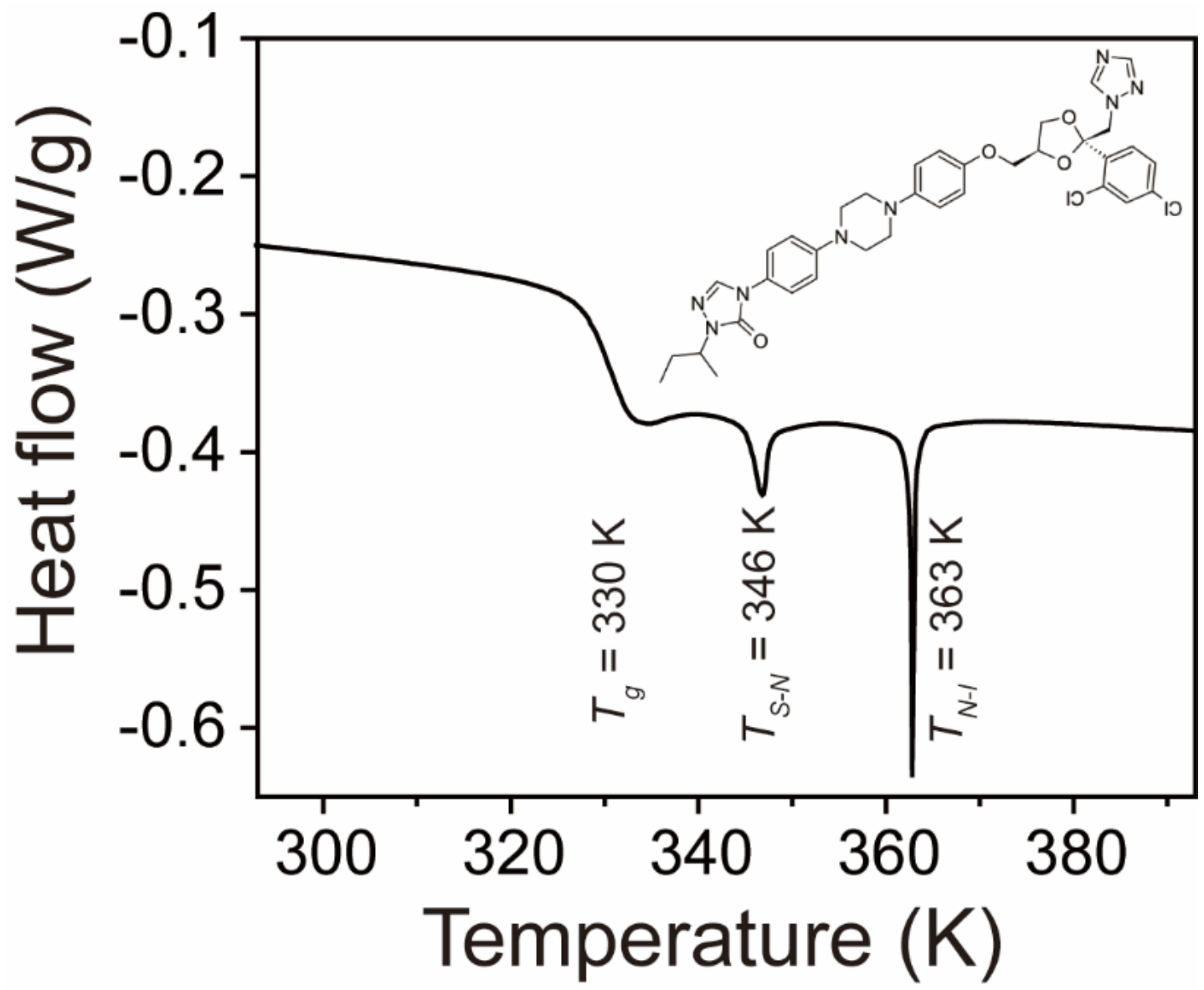

SI Figure 1 shows a DSC thermograph obtained upon heating a liquid-cooled itraconazole glass. There is an observed glass transition temperature at 330 K, as well as two liquid crystal transitions at 346 K and 363 K. Between 330 K and 346 K, itraconazole is in a smectic liquid crystalline phase, while between 346 K and 363 K, the system is in a nematic liquid crystalline phase. Above 363 K, the clearing point, itraconazole forms an isotropic liquid.

**SI Table 1. Incident angles used to obtain $\chi$ curves in Figure 4**

| $T_{substrate}$ (K) | Incident angle (degrees) |
|---|---|
| 337 | 2.960 |
| 334 | 2.960 |
| 331 | 2.960 |
| 329 | 2.974 |
| 327 | 2.998 |
| 324 | 3.013 |
| 322 | 3.044 |
| 319 | 3.093 |
| 317 | 3.144 |
| 314 | 3.228 |
| 312 | 3.324 |
| 309 | 3.403 |
| 307 | 3.481 |
| Annealed | 2.960 |

**SI Table 2. Table of integration limits for construction of $\chi$ curves in Figure 4**

| $T_{substrate}$ (K) | Integration limits, from $2\theta_{min}$ to $2\theta_{max}$ (degrees) |
|---|---|
| 337 | 5.4-6.4 |
| 334 | 5.4-6.4 |
| 331 | 5.4-6.4 |
| 329 | 5.6-6.6 |
| 327 | 5.6-6.6 |
| 324 | 5.6-6.6 |
| 322 | 5.6-6.8 |
| 319 | 5.65-5.9 |
| 317 | 5.7-7.0 |
| 314 | 5.9-7.2 |
| 312 | 6.0-7.2 |
| 309 | 6.0-7.2 |
| 307 | 6.0-7.2 |
| Annealed | 5.4-6.4 |

The two tables above are related to Figure 4 in the main text. These tables describe the geometry of the diffraction experiment used to obtain the data presented in the figure. SI Table 1 shows the incident angle used for every glass, in degrees. These angles were chosen to match the Bragg condition associated with the second order out-of-plane peak for the structures in the glasses. Table 2 shows the integration limits selected to obtain the 1D diffractogram plotted as a function of the azimuthal angle in the figure.

**SI Figure 2. Line cut along $q_{xy}$ at $q_z = 0$ for GIWAXS data on sample prepared at $T_{substrate} = 260$ K**

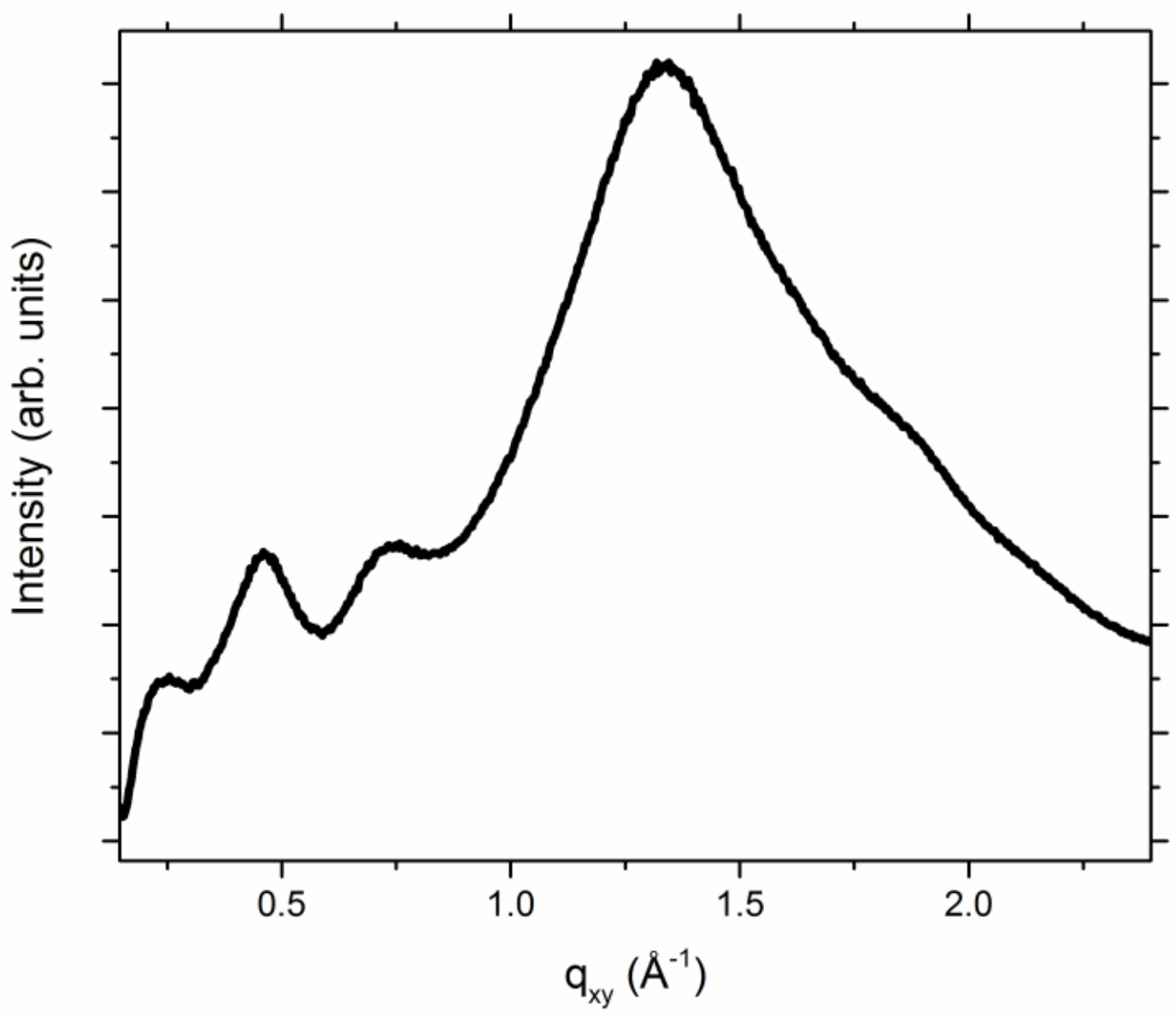

SI Figure 2 shows the 1D diffractogram of scattering intensity as a function of in-plane scattering vector $q_{xy}$. The in-plane diffraction pattern suggests that for the $T_{substrate} = 260$ K glass, there is a slight tendency for the itraconazole molecules to form in-plane smectic-like order, as indicated by the peaks observed at $q_{xy}$~0.24 Å$^{-1}$, ~0.48 Å$^{-1}$ and ~0.72 Å$^{-1}$. This may also be consistent with nematic-like order. Further investigation would be required to accurately describe the in-plane structure of the $T_{substrate} = 260$ K glass of itraconazole.

**SI Figure 3. Full-width-at-half-maximum of peaks along q**

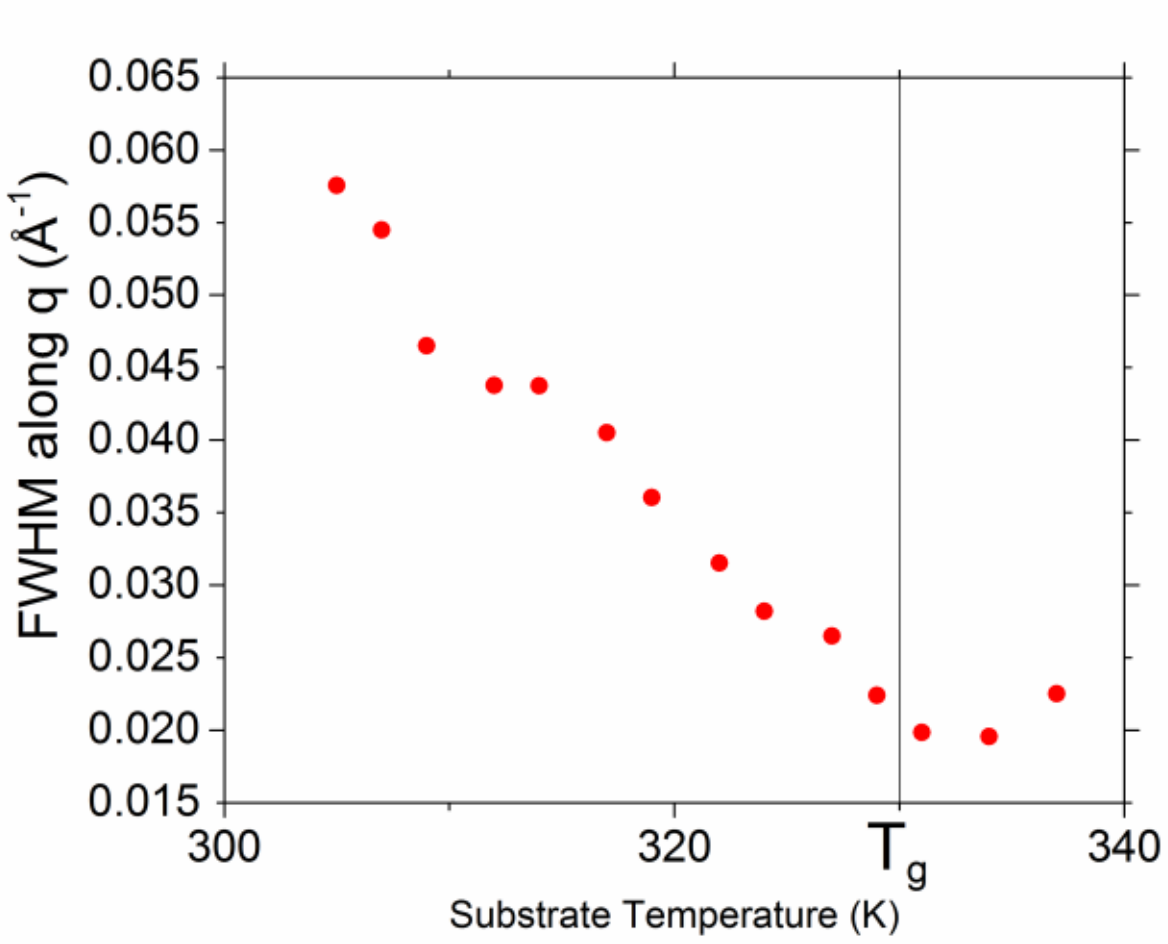

SI Figure 3 plots full-width-at-half-maximum (FWHM) along q (as defined in the top panel of Figure 2) as a function of $T_{substrate}$ of peaks shown in Figure 2. The broader the peak, the more disordered the layering structure associated with the peak. This broadening is related to the coherence length of the layered structure, as well as the layer spacing uniformity.